\documentclass{emulateapj}

\usepackage{amsmath}
\usepackage{natbib}
\usepackage{bm}
\usepackage[colorlinks=true,citecolor=blue,urlcolor=magenta,breaklinks]{hyperref}

\usepackage{graphicx}

\usepackage{cool}
\usepackage{eulervm}
\usepackage{times}

\usepackage{tikz,xcolor,hyperref}
\definecolor{lime}{HTML}{A6CE39}
\DeclareRobustCommand{\orcidicon}{%
	\begin{tikzpicture}
		\draw[lime, fill=lime] (0,0) 
		circle [radius=0.16] 
		node[white] {{\fontfamily{qag}\selectfont \tiny ID}};
		\draw[white, fill=white] (-0.0625,0.095) 
		circle [radius=0.007];
	\end{tikzpicture}
	\hspace{-2mm}
}
\newcommand{\orcidIL}{\href{https://orcid.org/0000-0002-5011-9195}{\orcidicon}}
 
\begin{document}
\title{The Sun: light dark matter and sterile neutrinos}
\author{Il\'idio Lopes$^1,^2$ \orcidIL}
\email[]{IL:ilidio.lopes@tecnico.ulisboa.pt}
\affiliation{$^1$Centro de Astrof\'{\i}sica e Gravita\c c\~ao  - CENTRA, \\
	Departamento de F\'{\i}sica, Instituto Superior T\'ecnico - IST,
	Universidade de Lisboa - UL, Av. Rovisco Pais 1, 1049-001 Lisboa, Portugal}
\affiliation{$^2$ Institut d'Astrophysique de Paris, UMR 7095 CNRS, Universit\'e Pierre et Marie Curie, 98 bis Boulevard Arago, Paris F-75014, France}

\date{\today}

\begin{abstract} 
Next-generation experiments allow for the possibility to testing the neutrino flavor oscillation model to very high levels of accuracy. Here, we explore the possibility that the dark matter in the current universe is made of two particles, a sterile neutrino and a very light dark matter particle. By using a 3+1  neutrino flavor oscillation model, we study how such a type of dark matter imprints the solar neutrino fluxes, spectra, and survival probabilities of electron neutrinos.
The current solar neutrino measurements 
allow us to define an upper limit for the ratio of the mass of a light dark matter particle $m_\phi$ and the Fermi constant  $G_\phi$, such that $G_\phi/m_\phi $ must be smaller than $10^{30}\; G_{\rm F} eV^{-1}$ to be in agreement with current solar neutrino data from the Borexino, Sudbury Neutrino Observatory, and Super-Kamiokande detectors. Moreover,  for models with a very small Fermi constant,  the  amplitude of the time variability must be lower than $3\%$  to be consistent with current solar neutrino data.  We also found that solar neutrino detectors like Darwin, able to measure neutrino fluxes in the low energy-range with high accuracy, will provide additional constraints to this class   of models that complement the ones obtained from the current solar neutrino detectors.
\end{abstract}

\keywords{
The Sun -- Solar neutrino problem --  Solar neutrinos --
Neutrino oscillations -- Neutrino telescopes -- Neutrino astronomy}

\maketitle

  
\section{Introduction}

The origin of dark matter has been a fundamental  problem in physics for almost six decades, during which most of the proposed solutions assumed that a single massive particle that interacts weakly with baryons makes all the dark matter observed in the universe \citep[e.g.,][]{2016RPPh...79i6901W}.
Recently, research has emerged where more sophisticated solutions have been proposed to solve the dark matter problem. One of these is the possibility of the dark matter being a composite of two light particles: a light dark matter (LDM)  particle $\phi$ and a sterile neutrino $\nu_s$. 

\medskip\noindent
The existence of such an LDM field can by identified with a dilation field of an extradimensional extension of the Standard Model or/and a CP-violating pseudo-Goldstone boson of a spontaneously broken global symmetry. For some of these models, $\phi$ couples to the Standard Model fields, and as such it induces periodic time variation in particle masses and couplings. In such theories the gauge invariance suggests that the $\phi$ should possess an identical coupling constant to charged leptons, in which case scalar interactions with the  electrons provide  a good opportunity for detection through  atomic clocks \citep[e.g.,][]{2016PhRvL.116c1102A}, accelerometers \citep[e.g.,][]{2018PhRvD..97g5020A}, and gravitational wave detectors \citep[e.g.,][]{2014ApJ...794...32L,2016PhRvD..93g5029G}. 

\medskip\noindent
Similarly, this  $\phi$ field can couple to neutrinos.
Once again, these types of interactions generically result in time-varying corrections to the neutrino masses, neutrino mass differences, and mixing angles, which can be searched for in the neutrino flux signals on present and future experimental neutrino detectors \citep{2013PhRvC..88b5501A, 2016PhRvD..94e2010A, 2018Natur.562..505B,2020arXiv200603114A}.
 If $\phi$ couples weakly to the neutrinos, over a large range of masses, it can  significantly modify the neutrino oscillations probabilities leading to a distorted survival electron neutrino probability function \citep{2016PhRvL.117w1801B, 2018PhRvD..97g5017K}.     

\medskip\noindent
The motivation for  such a model comes from the possibility of this composite particle physics model resolving two observational problems:


\smallskip\noindent
\begin{enumerate}

\item
The classical cold dark matter model leads to several inconsistencies with the cosmological 
observational data, such as the  missing satellite problem and the cusp problem \citep[e.g.,][]{2009NJPh...11j5029P}. Light dark matter resolves such problems if dark matter is totally or partially made of  light scalar particles with a mass of the order of $10^{-22} {\rm eV}$ \citep{2000PhRvL..85.1158H,2000ApJ...534L.127P}.
In hierarchical models of structure formation, such type of dark matter is able to explain the flatness observed on the profiles of the distribution of gas and stars in halos and filaments
\citep{2019PhRvL.123n1301M}.

\item
Although the standard three neutrino flavor model  produces a reasonable good global fit to all the neutrino data \citep{2019JHEP...01..106E}, there are now many hints that point out to the possibility of the existence of a fourth neutrino.  This one does not have any other interaction than gravity and for that reason is known as a sterile neutrino \citep{2019arXiv190600045D}. It was found that a flavor oscillation model made of the 3 active neutrinos plus a sterile neutrino could explain some of the observed  anomalies found on the Short baseline neutrino oscillation experiments \citep{2012PhRvD..86k3014G,2013PhRvD..87a3004G},  Liquid Scintillator Neutrino Detector \citep{2001PhRvD..64k2007A}, and  MiniBooNE Short-Baseline Neutrino Experiment \citep{2018arXiv180512028M}, as well as the  anomalies related with  GALLEX and SAGE solar-neutrino detectors -- the so-called  Gallium anomalies \citep{2019PhLB..795..542K}.
For instance, \citet{2019PhLB..795..542K} found that the data favour a $3+1$ neutrino
flavor model with $m_4 = \;1.1 eV$ and mixing matrix element $U_{e4} = 0.11$.
\end{enumerate}

\medskip\noindent
One possibility to resolve both problems (neutrino anomalies and structure formation) is to consider that dark matter is made of a light scalar field that couples to a sterile neutrino  \citep[e.g.,][]{2019PhLB..79734911F}.  The interactions of $\phi$ and $\nu_s$ could impede the oscillations in the universe and thereby improve the agreement between the structure formation and cosmological observations \citep[e.g.,][]{2014PhRvL.112c1803D,2014PhRvL.112c1802H}.

\medskip\noindent
If such $\phi$ and $\nu_s$ particles exist today, they were produced abundantly in the early universe. 
  For instance, sterile neutrinos can be produced via mixing with active neutrinos \citep{1994PhRvL..72...17D}, in some scenarios such neutrino production is being  enhanced by the oscillations between active and sterile neutrinos \citep{2019PhRvD..99h3507B,2020PhRvD.101j3516B,2020PhRvL.124h1802D}
or by the  lepton asymmetry \citep{1999PhRvL..82.2832S}. The production of light dark matter can take many forms, such as vector bosons by parametric resonance production \citep{2019PhRvD..99c5036D}. For instance, some models predict a  sterile neutrino abundance  of $ \Omega_s^2 h^2 =0.12 \left({\sin^2{(2\theta_s)}}/{3.5\times 10^{-9}}\right)
\left({m_{\nu_s}}/{7 keV}\right) $ where $m_{\nu_s}$ and  $\theta_s$ is the sterile neutrino  mass and mixing angle \citep{2009PhR...481....1K}. For the light dark matter field  some authors  obtained 
$ \Omega_\phi^2 =0.1 \left({a_o}/{10^{17}\; {\rm GeV}}\right)^2
\left({m_a}/{10^{-22}\; {\rm eV}}\right)^{1/2}$, where $a_o$ is a parameter that relates to the initial misalignment angle of the axion, and $m_a$ is the axion mass \citep{2017PhRvD..95d3541H,2019arXiv191207064N}. Conveniently, we will assume that in the present-day universe the  total dark matter abundance is given by
\begin{eqnarray}
\Omega_{\rm DM}h^2=\Omega_\phi^2 h^2 + \Omega_{\nu_s}^2 h^2 
\end{eqnarray}
where the $\Omega_\phi^2 h^2 $ and $\Omega_s^2 h^2 $ are the total   
$\phi$ and $\nu_s$  densities in the present universe, respectively.
For future reference, we assume that the present-day total dark matter abundance $\Omega_{\rm DM}h^2=0.12$ \citep{2018arXiv180706209P}, and the dark matter density  in the solar neighborhood is $\rho^\odot_{\rm DM}=0.39\; GeV \; cm^{-3} $ \citep{2010JCAP...08..004C}.

\medskip\noindent
In this paper, we study the impact that this light dark matter field has in the 3+1 neutrino flavor model. Specifically, we discuss how the light dark matter field modifies the neutrino flavor oscillations, and by using the current sets of solar neutrino data, we also put constraints in the parameters of such models and make predictions for the future neutrino experiments. 

\medskip\noindent
The article is organized as follows. In section \ref{sec-LDM}, we discuss how the  light dark matter drives the 3+1 neutrino flavor oscillations.   In section \ref{sec-LDMNM}, we present the neutrino flavor oscillation model in the presence of a cosmic light dark matter field. 
 In section \ref{sec-LDMeneutrinospectrum}, we compute the survival electron neutrino probabilities for the  electron neutrinos produced in the proton-proton
(PP) chain and carbon-nitrogen-oxygen (CNO) cycle solar nuclear reactions. In section \ref{sec:sun-neutrinos}, we discuss  the results in relation to current experiments and future ones.  
Finally, in  section \ref{sec-DC}, we present the conclusion and a summary of our results.  

\medskip\noindent   
If not stated otherwise,  we work in natural units in which $c=\hbar=1$. In these units all quantities are measured in GeV, and we make use of the conversion rules $1 m = 5.068 \times 10^{15} GeV^{-1}$, $1 kg = 5.610 \times 10^{26} GeV$ and $1 sec = 1.519 \times 10^{24} GeV^{-1}$. 

\section{Light Dark Matter and Sterile Neutrinos in the Universe}
\label{sec-LDM}

\medskip\noindent
We assume that in the present universe, the dark matter is composed of two fundamental particles: a light scalar boson $\phi$ and sterile neutrinos $\nu_s$, where  $m_\phi$ and  $m_{\nu_s}$ are their respective masses \citep{2014PhRvL.112c1802H}. The LDM field $\phi$ couples with the active neutrinos and the sterile neutrino by a  Yukawa interaction $g_\phi \phi \nu_s\nu_s$ where 
$g_\phi$ is a dimensionless coupling \citep{2019PhLB..79734911F}. To illustrate this effect, consider an LDM scalar $\phi$ with a Yukawa coupling to active neutrinos. Then the relevant part of the Lagrangian reads
\begin{eqnarray}
{\cal L} \supset -(m_{\nu} + g \phi)\nu\nu + H.c.,
\label{eq:Lagrangian} 
\end{eqnarray}
where for convenience of representation the flavor indices have been suppressed. We also assume  that the dimensionless coupling is very small  ($g \ll 1$). 
From the Euler-Lagrange equations of $\nu$ and $\phi$, it is possible to
show that the effect of $\phi$ on the  propagation of the neutrino is equivalent to changing the neutrino mass from $m_\nu$ to $m_\nu+\delta m_\nu$. As we will see later, this $\delta m_\nu$ perturbation will induce time variations in the mass-squared differences and mixing angles of all neutrino flavors through $\phi$. In principle the Yukawa couplings can have any structure in the neutrino flavor space.  In this work, we will focus on two convenient scenarios of great interest to neutrino detectors: mass-square differences and mixing angles \citep[e.g.,][]{2020arXiv200313448D}. Moreover, we will also assume that $\delta m_\nu=g_\phi \phi$ \citep[e.g.,][]{2019JHEP...12..046S}.

\medskip\noindent
\subsection{Dark Matter Time-dependent Variation}
\label{ssec-DMT}
The hypothesis that dark matter in the local universe is made of very light particles leads to the following description: the LDM field $\phi$ in the dark matter halo of the Milky Way is represented by a group of plane waves with frequency  $\omega_\phi$, such that $\omega_\phi=m_\phi(1+ v_\phi^2/2)$  where $v_\phi$ is the virial velocity of the particles in the dark halo. 
A population of such light particles will smooth inhomogeneities in the dark matter distribution on scales smaller than the de Broglie wavelength $\lambda_{\rm dB}$ of these LDM particles. For any particle, we compute $\lambda_{\rm dB}$ using the relation $\lambda_{\rm dB}=1.24\times 10^{22} \left(10^{-23}{\rm eV}/m_\phi\right)  \left(10^{-3}/v_\phi\right)\; {\rm cm} $. 

\medskip\noindent
We notice that the kinetic term on $\omega_\phi$ is neglected once the virial velocity  $v_\phi\sim 10^{-3}$ is very small \citep[e.g.,][]{2017PhRvL.118z1102B}.  Therefore, we dropped the corrections related with $v_\phi$ for the equation \ref{eq:Lagrangian}. Accordingly, the general form of this LDM field reads
\begin{eqnarray}
\phi(\vec{r},t)=\phi_o \cos{(m_\phi t+\epsilon_o)}
\approx \phi_o \cos{(m_\phi t)},
\label{eq:phi}
\end{eqnarray}
where $\phi_o$ and $\epsilon_o=\vec{v}_\phi\cdot\vec{r}$ are the amplitude and phase of 
the wave $\phi(t)$, respectively. In this work we consider $\epsilon_o\approx 0$.
Moreover, the energy momentum of a free massive oscillating field has a density given by $\rho_\phi=\phi_o m_\phi/2$  and a pressure
given by  $p_\phi=-\rho_\phi\cos{(2m_\phi t)}$. 
Although formally $\rho_\phi$ has an oscillating part proportional to $\phi(t)$, because this component is very small  we neglected its contribution in this analysis  \citep{2014JCAP...02..019K}.     

\medskip\noindent

The quantity  $\phi_o$ is a slowly varying function of the position. 
Conveniently, the amplitude of $\phi(t)$ can be written as $\phi_o=\sqrt{2\rho_{\phi}(r)}/m_\phi$ where $\rho_{\phi} (r)= \rho^\odot_{\rm DM}\left(\Omega_{\phi}/\Omega_{\rm DM}\right)$ is the fraction of dark matter density in $\phi$ particles at the space-time coordinate $\vec{r}$. Accordingly,  the Sun immersed in this light dark matter halo will experience a periodic perturbation due to the action of the $\phi(\vec{r},t)$, which by the presence of a Yukawa coupling $g_\phi$ will exert a temporal variation on the propagation of all neutrinos.  
We estimate the dark matter density number $n_\phi $ in the solar neighborhood as follows: if we  consider that the main contribution arises from a single dark matter particle with mass $m_\phi$, then the relevant density in our case will take the value $n_\phi=\rho_\phi/m_\phi$. If we assume that all dark matter is made of $\phi$ bosons,  we have $\rho_\phi=\rho^\odot_{\rm DM}=0.39\; GeV \; cm^{-3} $
\citep{2010JCAP...08..004C}
 and $m_\phi= 10^{-22}\; {\rm eV}$ then $n_\phi=3.9\times 10^{30}\;cm^{-3} $ (particles per centimeter cubed). This value is only 2 orders of magnitude smaller than the density of electrons in the Sun's core, $n_e\sim 6\;10^{31}\;cm^{-3}$ \citep{2013ApJ...765...14L}.  Since these particles are very light, we assume that there is no accretion of these particles in the Sun's core during its evolution in the main sequence until the present age. 
 
 \medskip\noindent
\subsection{Neutrino Time-dependent  Dark-matter-induced Oscillations}

In the presence of the LDM field $\phi$, the neutrino mass $m_\nu$, according to equation \ref{eq:Lagrangian} \citep{2020arXiv200313448D}, will  receive a contribution 
$\delta m_\nu=g\phi$, such that from equation \ref{eq:phi}, we obtain
\begin{eqnarray}
\frac{\delta m_\nu}{m_\nu} =\epsilon_\phi \cos{(m_\phi t)},
\end{eqnarray}
where $\epsilon_\phi$ is the amplitude 
\begin{eqnarray}
\epsilon_\phi=
\frac{g_\phi\sqrt{2\rho_{\phi}}}
{m_\phi m_{\nu}}=
\frac{g_\phi\sqrt{2\rho^{\odot}_{\rm DM}}}
{m_\phi m_{\nu}}\; \left(\frac{\Omega_{\phi}}{\Omega_{\rm DM}}\right)^{1/2}.
\label{eq:etaphi}
\end{eqnarray}
If not stated otherwise, we will assume that all dark matter in the present universe is made of only LDM particles such that $\Omega_{\phi}=\Omega_{\rm DM}$.
We observe that $\epsilon_\phi $ is a relevant factor even if $\phi$  is a small fraction of the dark matter halo.  In particular, $\phi$ will affect the oscillation parameters of all neutrino flavors, including the sterile  sterline neutrinos. 
 If we only take into account the first order perturbation, thus, the neutrino mass-squared difference can be written as  
\begin{eqnarray}
\Delta m_{ij}^2(t)=m_i^2-m_j^2\approx 
\Delta m^2_{ij,o}[1+2\epsilon_\phi\cos{(m_\phi t)}] 
\label{eq:deltamphi}
\end{eqnarray}
where $\Delta m^2_{ij,o}$ is the standard (undistorted) value and $\Delta m^2(t)$ evolves through  $\phi(t)$ (see Equation \ref{eq:phi}), with an amplitude $\epsilon_\phi $ (see Equation \ref{eq:etaphi}), and a frequency $m_\phi$.  The mass-squared difference $\Delta m^2_{ij}$ 
 between neutrinos of different flavors follows the usual convection \citep[e.g.,][]{2017PhRvD..95a5023L} such  that $\Delta m^2_{i1}=m^2_{i}-m^2_{1}$ ($i=$2, $3$,  $4$). In particular for the sterile neutrino,
 we have  $\Delta m^2_{41}= m^2_{4}-m^2_{1}$ where $m_4$ is the mass of the sterile neutrino.
Similarly, the mixing angles variation is written as
\begin{eqnarray}
\theta_{ij}(t)\approx 
\theta_{ij,o}+\epsilon_\phi\cos{(m_\phi t)},
\label{eq:thetaphi}
\end{eqnarray} 
where $\theta_{ij,o}$ is the standard (undistorted) mixing angle. The indexes 
$i$ and $j$ in $\theta_{ij}$ follow a convention identical but not equal for the mass-squared differences \citep[see ][and references therein]{2018ApJ...869..112L}.  Therefore, as first suggested by 
\citet{2018PhRvD..97g5017K}, the LDM $\phi(t)$ impacts the neutrino flavor oscillations through the modified expressions for the mass-squared differences (Equation \ref{eq:deltamphi})  and mixing angles (Equation  \ref{eq:thetaphi}).

\section{Light Dark Matter and the Sterile Neutrino Model} 
\label{sec-LDMNM}
In the following section, we consider a 3+1 neutrino flavor oscillation model to describe
the propagation of active neutrinos ($\nu_e$, $\nu_{\tau}$, $\nu_{\mu}$) plus a sterile neutrino $\nu_s$  through the solar plasma.  Following the usual notation
$(\nu_e,\nu_\tau,\nu_\mu,\nu_s)$ corresponds to the neutrino  flavors,
($\nu_1$, $\nu_{2}$, $\nu_{3}$, $\nu_4$)  are the mass neutrino eigenstates,
and ($m_1$, $m_{2}$, $m_{3}$, $m_4$) are the neutrino masses.
 The evolution of  neutrinos propagating in  matter is described by the see 
\begin{eqnarray}
i\frac{d\Psi}{dr} ={\cal H}\Psi =\frac{1}{2E} \left( \mathbf{U}M^2\mathbf{U}^\dagger +2E{\cal V} \right)\Psi, 
\label{eq:Hamiltonian1}
\end{eqnarray}
where ${\cal H}$ is the Hamiltonian and $\Psi=(\nu_e,\nu_\tau,\nu_\mu,\nu_s)^{T}$.  $M^2$ is  a neutrino mass matrix, $\mathbf{U}$ is a ($4\times 4$) unitary matrix describing the mixing of neutrinos  and
${\cal V} $ is the diagonal matrix of Wolfenstein potentials \citep{1989RvMP...61..937K}. 
 $M^2$  is defined as
$M^2=\mathrm{{ diag }}\{0,\Delta m^2_{21},\Delta m^2_{31},\Delta m^2_{41}\}$.
The first term of the Hamiltonian describes the neutrino propagation through vacuum  and the second term incorporates the matter effects or  Mikheyev-Smirnov-Wolfenstein
(MSW) effects \citep[][]{1978PhRvD..17.2369W,1985YaFiz..42.1441M}. In general, the Hamiltonian ${\cal H}$ that drives the evolution of neutrino flavor must include the  Wolfenstein potentials related with $\phi(t)$ \citep{2018PhRvD..97d3001B}.   

\medskip\noindent
In most studies of three-neutrino flavor models, the authors are solely interested in the modulation coming from the square mass differences $\Delta m^2_{ij}(t)$ (by Equation \ref{eq:deltamphi}) and mixing angles $\theta_{ij}(t)$ (by Equation  \ref{eq:thetaphi}). For that reason, all neutrinos are assumed to couple $\phi(t)$. As a consequence, their contribution to ${\cal V} $ cancels out. Hence, it is correct to neglect the contribution of $\phi(t)$ to the Wolfenstein potential \citep{2020arXiv200703590D}. Nevertheless, here in this 3+1 neutrino flavor model, as we will discuss later, we include the contribution of  $\phi(t)$  in ${\cal V} $.

\medskip\noindent
This  3+1 neutrino flavor model with dark matter is identical to the standard  (undistorted)  three-neutrino flavor model (see Equation \ref{eq:Hamiltonian1}). However, in this model we included a sterile neutrino, and the Wolfenstein potentials in ${\cal V}$  are modified to take into account the new LDM field $\phi$ \citep{2015PhLB..744...55M}.

\medskip\noindent
\subsection{Neutrino Matter-induced Oscillations}   
\begin{figure}[!t] 
\centering 
\includegraphics[scale=0.47]{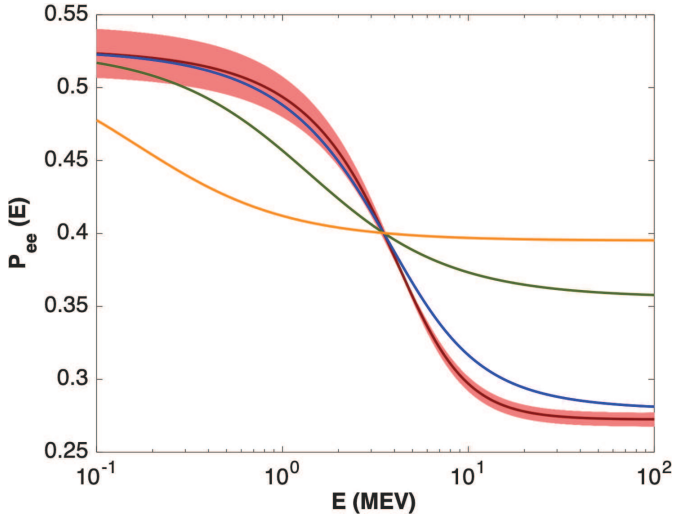}	\caption{The survival probability electron neutrinos $P_{ee}(E,\phi)$ and $\langle P_{ee}(E)\rangle$  in a  3+1  neutrino flavor oscillation model in which neutrino couple to the light dark matter field  $\phi$ with 
fixed value of  $m_\phi$  and $\epsilon_\phi$ (see main text). The figure shows  $\langle P_{ee}(E)\rangle$ (with $\epsilon_\phi \approx 0\%$) for the  following LDM  models  with $G_\phi/m_\phi$ ratios:  $10^{29}\; {\rm G_{\rm F}\, eV^{-1}}$ (blue curve),  
$10^{30}\; {\rm G_{\rm F}\, eV^{-1}}$ (green curve),  $10^{31}\; {\rm G_{\rm F}\, eV^{-1}}$ (orange curve).   
The figure also shows a model with $G_\phi\approx 0$ (red curve) and 
an ensemble of $P_{ee}(E,\phi)$ corresponding to different time-varying mass-square differences and angles for  $\epsilon_\phi=1.5\%$ (pink band). See main text for details.}
	\label{fig:Petime1}
\end{figure}

\medskip\noindent
In the standard three-neutrino flavor model\footnote{In this  model, the intermediate particle is an heavy boson, specifically  the $Z$ or $W^{\pm}$ bosons.}, the matter potential  ${\cal V}$ takes into account the interaction of active neutrinos ($\nu_e,\nu_\mu,\nu_\tau$) with the ordinary fermions  of the solar plasma,  for which the  
${\cal V}=\mathrm{{diag }}\{V_{cc}+V_{\rm nc},V_{\rm nc},V_{\rm nc}\}$
where $V_{\rm cc}$ corresponds to the weak charged current ($\rm cc$) that takes into account the
forward scattering of $\nu_e$ with electrons, and  $V_{\rm nc}$ is the weak neutral current ($\rm nc$) 
that corresponds to the scattering of the active neutrinos with the ordinary fermions of the solar plasma \citep[e.g.,][]{2020PhR...854....1X}. 
 $V_{\rm nc}$ can be expressed as $V_{\rm nc}=V^e_{\rm nc}+V^p_{\rm nc}+V^n_{\rm nc}$ where $V^j_{\rm nc}$
 with $j=$ $e$, $p$, $n$ are the contributions coming from electrons, protons, and neutrons, respectively. 
However due to the electrical neutrality of the solar plasma, the contribution of $V^e_{\rm nc}$  
and $V^p_{\rm nc}$ canceled out such that $V_{\rm nc}=V^n_{\rm nc}$.
Accordingly, $V_{\rm cc}= \sqrt{2}G_{\rm F}\; n_e(r)$ and $V_{\rm nc}=V^n_{\rm nc}=G_{\rm F}/\sqrt{2}\;n_n(r)$.
Here $G_{\rm F}$ is the Fermi constant and $n_e(r)$ and $n_n(r)$
 are the number density of electrons and neutrons inside the Sun.
 Nevertheless, since  $V_{\rm nc}$ is an universal term for all active neutrino flavors, and as such does not change the flavor oscillations pattern,
 conveniently we write ${\cal V}=\mathrm{{diag }}\{V_{\rm cc}+0,0,0\}$.
 Now, the inclusion of sterile neutrinos in the neutrino flavor model alters ${\cal V}$ (from Equation \ref{eq:Hamiltonian1}) by incorporating  a new degree of freedom, as a consequence 
${\cal V}=\mathrm{{ diag }}\{V_{\rm cc}+V_{\rm nc},V_{\rm nc},V_{\rm nc},0\}$ \citep[e.g.,][]{ 2009PhRvD..80k3007G,2016EPJA...52...87M,2020PhR...854....1X}.

\medskip\noindent
Finally, in our 3+1 neutrino flavor model, we include the interaction of active and sterile neutrinos with the dark matter field $\phi$ by means of an intermediate heavy boson $I$\footnote{
We assume the boson $I$ has a mass $m_I$ identical to $Z$ and $W^{\pm}$ bosons.}. These interactions result from the forward scattering of these neutrinos through the LDM field $\phi$, thus ${\cal V}=\mathrm{{diag }}\{V_{\rm cc}+V_{\rm nc}+V_{\nu_e\phi},V_{\rm nc}+V_{\nu_\mu\phi},V_{\rm nc}+V_{\nu_\tau\phi},V_{\rm nc} +V_{\nu_s\phi} \}$,
where $V_{\nu_i\phi}$ (with $\nu_i=\nu_e,\nu_\mu,\nu_\tau,\nu_s$)  relates to the neutrino $\nu_i$.
This ${\cal V}$ corresponds to a generalization of the Wolfenstein potentials found in the literature, for which most neutrino flavor  models only take into account the scattering of the sterile neutrinos on  heavy dark matter  \citep{2017JCAP...07..021C,2018ApJ...869..112L,2019PhRvD..99b3008L}.

\medskip\noindent
In our model,  we opt to assume that all active neutrinos experience the same interaction with the LDM field $\phi$, such that their dark matter potentials are the same, such that   $V_{\nu_j\phi}=V_{\nu_a\phi}$ (with $j=e,\mu,\tau$), it follows 
${\cal V}=\mathrm{{diag }}\{V_{\rm cc}+V_{\rm nc}+V_{\nu_a\phi},V_{\rm nc}+V_{\nu_a\phi},V_{\rm nc}+V_{\nu_a\phi},V_{\rm nc} +V_{\nu_s\phi} \}$. Now, if we subtract the common term $V_{\rm nc}+V_{\nu_a\phi}$ to the diagonal matrix ${\cal V}$, the latter takes 
the simple form: ${\cal V}=\mathrm{{diag }}\{V_{\rm cc},0,0,V_{\nu_s\phi}-V_{\nu_a\phi}-V_{\rm nc}\}$.

\medskip\noindent
The potential $V_{\nu_i\phi}$ (with $i=a,s$)  is given by $V_{\nu_i\phi}=G_{\nu_i\phi} n_{\phi}$  where $G_{\nu_i\phi}$ is the equivalent of the Fermi constant and $n_{\phi}$ is the distribution of dark matter inside the Sun \citep{2019JHEP...12..046S}. 
Equally,  $V_{\nu_i\phi}$ relates directly with the local density of dark matter  $\rho^\odot_{\rm DM}$ by the expression: $ V_{\nu_s\phi}= \left(G_{\nu_s\phi}/m_\phi\right)  \left( \rho^\odot_{\rm DM} \Omega_\phi/\Omega_{\rm DM}\right)$,  where 
we assume  the ratio $G_{\nu_s\phi}/m_\phi$ is a free parameter of the LDM model.
The generalized Fermi constant is defined as $G_{\nu_i\phi}=g_{\nu_i}g_\phi /m_I^2$ where   $g_{\nu_i} $  represents the coupling constant of the corresponding neutrino $\nu_i$,  
and $m_I$ is the mass of the intermediate boson $I$  \citep{2015PhLB..744...55M}. 
This expression for the potential $V_{\nu_i\phi}$ is valid since we assume  that $m_I^{-1}\ll R_\odot$ where $R_\odot$ is the solar radius\footnote{As an example, if we consider $m_I\approx m_Z \approx 90 GeV$ where $m_Z$ is the mass of the $Z$ boson, then the propagation of neutrinos (like of the neutral current) verifies the condition $m_I^{-1}\ll R_\odot$.} \citep{2019JHEP...12..046S}. In general, we could expect that the contribution of $\phi(t)$ to $V_{\nu_i\phi}$ could lead to a time-dependent relation, however, as discussed previously (in section \ref{ssec-DMT}) and mentioned for the first time by \citet{2014JCAP...02..019K}, this is because the oscillatory component on the local density relates with $v_\phi^2$. This term is minimal, and therefore we neglected it.

\medskip\noindent
In this  preliminary study, without loss of generality, we choose to simplify 
${\cal V}$ further: since the term $V_{\nu_s\phi}-V_{\nu_a\phi}-V_{\rm nc}$  has  two Wolfenstein potentials ( $V_{\nu_a\phi}$ and  $V_{\nu_s\phi}$)
that effectively correspond to two new degrees of freedom,
both of these have  an identical impact on the neutrino flavor oscillation model.  We choose to simplify the model by assuming that $V_{\nu_a\phi}$   is much smaller than $V_{\rm nc}$. Consequently,  ${\cal V}$  takes the simplified form:  ${\cal V}\approx\mathrm{{diag }}\{V_{\rm cc},0,0,V_{\nu_s\phi}-V_{\rm nc}\}$.  For reference, we note that in the Sun's core  $V_{\rm nc}$ is  always smaller than $V_{\rm cc}$, once $n_e$ is more than twice as lager as $n_n$
\citep[e.g.][]{2018EPJC...78..327L}.  This potential is identical to others found in the literature, for instance in \citet{2017JCAP...07..021C} and \citet{2018ApJ...869..112L}.   Therefore, the matter potential $V_{\nu_s\phi}$  reads $ V_{\nu_s\phi}= G_{\nu_s\phi} n_\phi$ where  for convenience of analysis, we choose to define  the generalized Fermi constant as  $G_{\nu_s\phi}= 4\sqrt{2}G_\phi G_{\rm F} $  where $G_{\phi}$ is our free parameter.  Since these dark matter particles have a mass much smaller than $4\; {\rm GeV}$, the solar plasma conditions do not allow the accretion of dark matter by the Sun \citep[e.g.,][]{2019ApJ...879...50L}, therefore we will assume that  the distribution of dark matter inside the star is equal to the value measured for the solar neighborhood $n_{\phi}$ (see section \ref{ssec-DMT}).

\begin{figure}[!t]
	\centering 
	\includegraphics[scale=0.45]{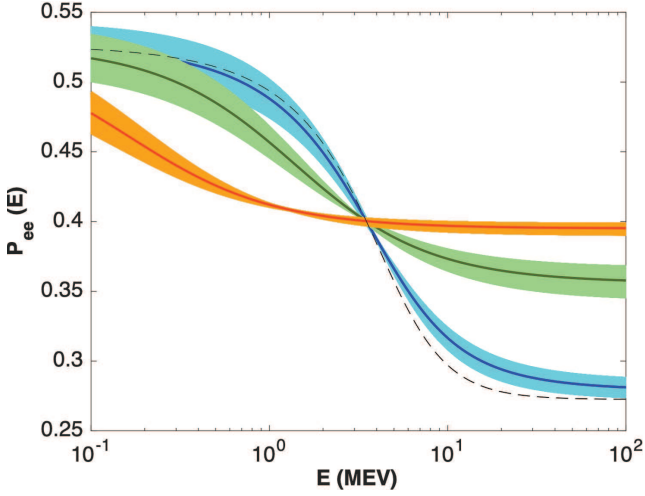}
	\caption{The survival probability electron neutrinos $P_{ee}(E,\phi)$ and the corresponding $\langle P_{ee}(E)\rangle$  for some of the models shown in figure \ref{fig:Petime1} for which $\epsilon_\phi=1.5\%$ :  $G_\phi=10^{20}\; G_{\rm F}$ (blue band and dark blue curve), $G_\phi=10^{21}\; G_{\rm F}$ (green band and dark green curve) and  $G_\phi=10^{22}\; G_{\rm F}$ (orange band and dark red curve). The figure also shows $\langle P_{ee}(E)\rangle$  for  the 3+1 model with no time dependence ($G_\chi\approx 0$)  and $\epsilon_\phi\approx 0$) as a thin dashed black curve. The latter curve corresponds to the red curve in figure \ref{fig:Petime1}.}
	\label{fig:Petime2}
\end{figure}

\subsection{Neutrino Flavor Oscillation Model and the Survival Probability of Electron Neutrinos}   

If we adopt as reference the current experimental set of parameters for the active neutrinos \citep[e.g.,][]{2019JHEP...01..106E}, the propagation neutrinos in the solar interior are completely adiabatic. The same is valid for the 3+1 neutrino flavor oscillation model coupled to an LDM field $\phi$ considered in this study.  
Conveniently, the propagation of neutrinos away from resonances 
is well represented by a two neutrino flavor oscillation
model.  The motivation for such approximation can be found in \citet{2018EPJC...78..327L} and references therein.  In such a case, the electron neutrino flavor oscillation is dominated by the  $(\nu_1,\nu_2)$ mass eigenstates and is only slightly affected by the decoupled   $(\nu_3,\nu_4)$  eigenstates, since the associated mixing angles for the latter pair are very small \citep{1986PhRvL..57.1805K}. 
Moreover,  $\nu_3$ and $\nu_4$  evolve independent of each other and are completely independent of the doublet  $(\nu_1,\nu_2)$.  
In this limit, as proposed by several authors \citep[e.g,][]{2011PhRvD..83k3013P,2013arXiv1306.2903B}, 
the split of the 3+1  neutrino flavor model into a dominant two neutrino flavor model ($\nu_e,\nu_\mu$) with additional corrections for  $\nu_\tau$ and $\nu_s$ significantly simplified  the calculation and allowed us to obtain an analytical solution \citep[e.g.,][]{1989RvMP...61..937K}.    
 
\medskip\noindent
Among the many expressions available in the literature  to compute  the  survival probability of electron neutrinos $P_{e}$~\citep[e.g.,][]{2000NuPhB.583..260L,2015PhLB..744...55M}
in a 3+1 neutrino flavor model developed in the approximate scenario of a two-flavor neutrino model~\citep[e.g.,][]{1989RvMP...61..937K}, we opted to choose the expression obtained  by~\citet{2017JCAP...07..021C}
for the case in which $V_{\rm cc}E/\Delta m_{31}^2 \ll 1$ (and $s_{34}=0$) which has a better numerical accuracy than others. In that case  the survival probability of electron neutrinos, i.e.,  $P_{e}$ $[\equiv P (\nu_e \rightarrow \nu_e]]$, reads
\begin{eqnarray}
P_{ee}(E,\phi)=s_{13}^4+c_{13}^4c_{24}^4c_{14}^4
+a_m+b_m,
\label{eq:Peesunrad}
\end{eqnarray} 
where $c_{ij}=\cos{\theta_{ij}}$ and $s_{ij}=\sin{\theta_{ij}}$.
The functions  $a_m$ and $b_m$ are dependent on
the internal structure of the Sun and are given by the expressions: 
\begin{eqnarray}
a_m= C_1 (s_{m}c_{14}-c_{m}s_{14}s_{24})^2
\end{eqnarray}
and
\begin{eqnarray}
b_m= C_2 (c_{m}c_{14}+s_{m}s_{14}s_{24})^2
\end{eqnarray}
where $c_m=\cos{\theta_m}$, $s_m=\sin{\theta_m}$, 
$C_1=c_{13}^4(c_{14}s_{12}-c_{12}s_{14}s_{24})^2$ and 
 $C_2=c_{13}^4(c_{12}c_{14}+s_{12}s_{14}s_{24})^2$.
The angle $\theta_{m}$ is obtained for the present-day Sun
\citep[i.e., the standard solar model, see details of this model in ][]{2013MNRAS.435.2109L} using  the expression~\citep{2017JCAP...07..021C}:
$\cos{(2\theta_{m})}=
{\cal M }_x({\cal M }_y^2+{\cal M }_x^2)^{-1/2},$
where ${\cal M }_x\equiv \cos{(2\theta_{12})} -\eta_{\nu}V_x$ and ${\cal M}_y\equiv |\sin{(2\theta_{12})} +\eta_{\nu}V_y|$. $\eta_{\nu}$ is the ratio of the energy of the neutrino $E$ in relation to  $\Delta m^2_{21}$ given by $\eta_{\nu} (E)=4E/\Delta m_{21}^2$.  The functions $V_x$ and 
$V_y$ are given by
\begin{eqnarray}
V_x=\frac{1}{2}\left[V_{\rm cc} c^2_{13}
(c^2_{14}-s_{14}^2s_{24}^2) + V_s
(s_{14}^2-c_{14}^2s_{24}^2) \right],
\end{eqnarray}
\begin{eqnarray}
V_y=(V_s-V_{\rm cc}c_{13}^2)c_{14}s_{14}s_{24}
\end{eqnarray}
and
\begin{eqnarray}
V_{s}=V_{\nu_s\phi}-V_{\rm nc}.
\label{eq:Vs}
\end{eqnarray}

\begin{figure}[!t]
	\centering 
	\includegraphics[scale=0.45]{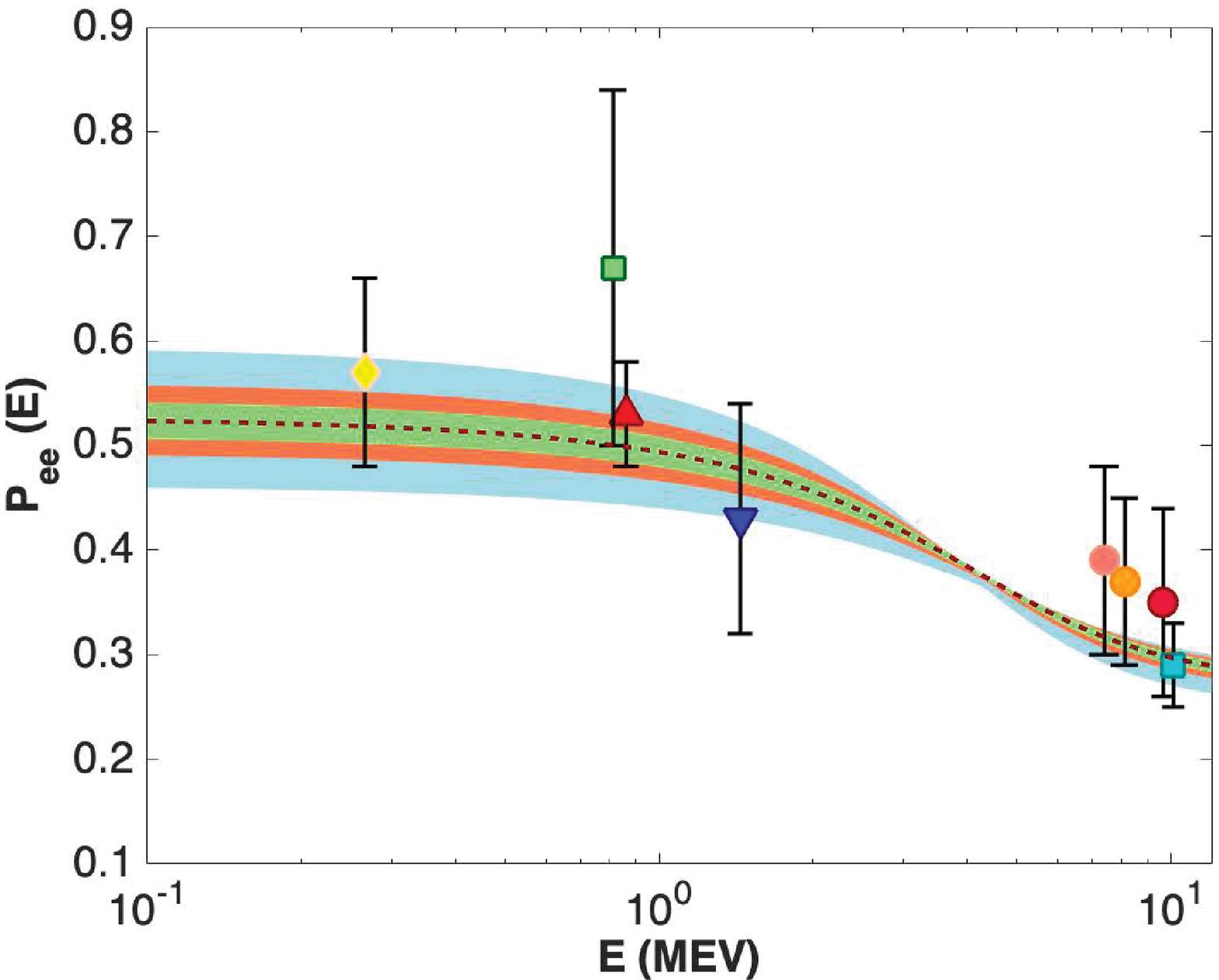}
	\caption{The survival probability of the electron neutrinos for several solar nuclear reactions: $pp$ (yellow diamond),  $^7Be$ (green square), $^7$Be (red upward triangle), $pep$ (blue downward triangle) and  $^8$B HER (salmon circle), $^8$B HER-I (orange circle), $^8B$ HER-II (magenta circle) and $^8B$ (cyan square). The diamond, triangle, and circles points use data from the Borexino \citep[][]{2020PhRvD.101f2001A,2019PhRvD.100h2004A,2018Natur.562..505B,2010PhRvD..82c3006B}  and the two square data points correspond to data obtained from SNO \citep{2013PhRvC..88b5501A} and Super-Kamiokande  
		\citep{ 2016PhRvD..94e2010A,2008PhRvD..78c2002C} detectors.  Each color band corresponds  to an ensemble of $P_{ee}(E,\phi)$ (Equation \ref{eq:Peesunrad})
		in a 3+1 neutrino model with $G_\phi\approx 0$ and  $\epsilon_\phi\approx 0$ (dashed curve), 1.5\% (green), 3.0\% (orange) and 6.0\% (light blue). These survival probabilities  are computed using an high-Z SSM (see the main text) and the experimental values from the
		different neutrino experiments.}
	\label{fig:Peeidata1}
\end{figure}

\subsection{Light Dark Matter Impact on Solar Neutrinos}
\label{sec-LDMeneutrinos}

The  survival probability of electron neutrinos (Equation \ref{eq:Peesunrad}) is a time-dependent  function through  equations  (\ref{eq:deltamphi}), (\ref{eq:thetaphi}) and (\ref{eq:phi}).  Conveniently  we define an effective oscillation probability $\langle P_{ee}(E) \rangle$ that corresponds to an ensemble average of all the $ P_{ee,}(E,\phi)$ (Equation \ref{eq:Peesunrad}),  as such 
 \begin{eqnarray}
 \langle P_{ee}(E) \rangle=\int_0^{\tau_\phi} P_{ee} (E,\phi)
 \frac{dt}{\tau_\phi},
 \label{eq:Peet}
 \end{eqnarray}
 where   $\tau_\phi=2\pi/m_\phi$ is the period of the LDM field $\phi(t)$. 

\medskip\noindent
The ability of a solar neutrino detector to measure the impact of the time-dependent LDM field $\phi(t)$ on the survival probability  $P_{ee}(E,\phi)$ (Equation \ref{eq:Peesunrad}) depends on three characteristic time scales:
the neutrino flight time $\tau_\nu$, the time between two consecutive neutrino detections  $\tau_{ev}$, and  the total run time of the experiment  $\tau_{ex}$. The neutrino flight time is proportional to the Earth-Sun distance $d_{\earth}$ such that $\tau_\nu=d_{\earth}/c\approx 8.2\; {\rm min}$ where $c$ is the speed of light.  The number of events measured by a detector varies strongly from one to another.

\medskip\noindent
The next generation of experiments will have $\tau_{ev}$  much larger than the pioneer Homestake experiment that only detects a few events per year \citep{1976Sci...191..264B}. The forthcoming Jiangmen Underground Neutrino Observatory  \citep[JUNO;][]{2015arXiv150807166A}  experiment expects to measure a few tens of neutrinos per day (for instance 200 events per day or $\tau_{ev}\approx {\rm 7\; min}$). The total experimental run time for most solar neutrino detectors is of the order of a few decades (for instance $\tau_{ex}\approx {\rm 10\; yr}$), and future experiments will also have significant running times.  Hence for all models considered in this study, we assume that solar neutrino detectors will run for long periods and will collect a large number of events, therefore we  assume that $\tau_{ev}$ and $\tau_{ex}$ have sufficient 
small and large values, respectively. 

\medskip\noindent
In such conditions, the solar neutrino spectra time modulation by $\phi(t)$ depends on the period $\tau_\phi$ of the LDM field  in comparison to the flight time of solar neutrinos $\tau_\nu$.  Since these neutrinos have a $\tau_\nu\approx 8.2\; {\rm min}$, it is possible to find the  value of $m_\phi$  for which  $\tau_\nu=\tau_\phi$ which occurs for $m_{\phi,c}=8.3\; 10^{-18}\;{\rm eV}$. 
Accordingly, we can define two regimes for the time modulation of survival probability of electron neutrinos : 

\smallskip\noindent 
\begin{enumerate}
\item	
For $ \tau_{\phi} \ge \tau_{\nu} $ (low-frequency regime or low LDM mass),
the time modulation of  $ P_{ee} (E,\phi)$ occurs when the period of $\phi(t)$ is  larger than $\tau_{\nu}$. In this case a temporal variation of the neutrino signal may be observed.
This corresponds to a LDM field with a mass such that $m_\phi \le m_{\phi,c} $. Therefore, the  LDM field can induce an observable time variation in neutrino oscillation measurements as periodicity  in the  solar neutrino fluxes \citep{2016PhRvL.117w1801B}. Obviously, if $ \tau_{\phi}$ becomes very large, the modulation of  $ P_{ee} (E,\phi)$ becomes indistinguishable from the standard  scenario (undistorted case), since the running time of the experiment is not sufficient  to observe this phenomena.
Nevertheless, in our study, the LDM field has always an $m_{\phi}\ge  10^{-23}\;{\rm eV}$  or a period  $ \tau_{\phi} \ge 13\;{\rm yr} $. Therefore, it is always possible to probe such a model with current experimental running times.

\item
For $ \tau_{\phi} \le \tau_{\nu} $ (high-frequency regime or high LDM mass), the  change of $ P_{ee} (E,\phi)$  due to $\phi(t)$ is too fast to be observed as a modulating signal like in the previous case.  This regime occurs for LDM fields with a mass such that  $m_\phi \ge m_{\phi,c} $. 
Nevertheless, the time average of the ensemble of oscillation probability $P_{ee}(E,\phi)$ can be distorted in such a regime, hence the effect can be detected as $\langle P_{ee}(E) \rangle$ which will deviate from the standard scenario \citep{2018PhRvD..97g5017K}.  
The net effect of averaging over time induces a shift in the observed values of $ P_{ee} (E,\phi)$ relative to its undistorted value.
\end{enumerate}

\smallskip\noindent 
Therefore, we can expect to study both regimes in a quite reasonable range of LDM masses using data from the present and future solar neutrino experiments. In fact, some of the current solar neutrino detectors have already large statistics and high event rates that we can use to look for time modulations in solar neutrinos. Some of these neutrino collaborations have already searched for regular  phenomena with periods varying  from 10 minutes to 10 yr \citep[e.g.,][]{2003PhRvD..68i2002Y,2010ApJ...710..540A}.
  
\medskip\noindent
In this work, we will study models that will fall in these two regimes of time modulation. Therefore to satisfy the conditions mentioned above, we decided to analyze the impact of the LDM field in solar neutrino fluxes for $\phi(t)$ with a period $\tau_\phi$ varying from $4\;\mu {\rm s}$ to $13\;{\rm yr}$ or equivalently with a $m_\phi$ varying  from $10^{-9}\;{\rm eV}$ to $10^{-23}\;{\rm eV}$, which is a range possible to be scanned by future detectors like Deep Underground Neutrino
Experiment \citep{2015arXiv151206148D} and JUNO \citep{2016JPhG...43c0401A}.

 \begin{figure}[!t]
 \centering 
 \includegraphics[scale=0.45]{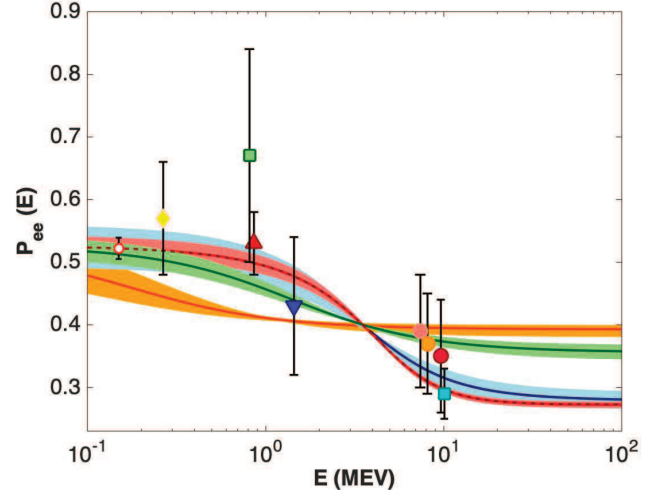} 
 \caption{The survival probability of the electron neutrinos for  several LDM models
 (see the main text and caption of figure \ref{fig:Peeidata1} and \ref{fig:Petime2} for details).  The LDM models are identical to the ones presented in Figure \ref{fig:Petime2}, with the following changes: (i) $G_\phi=0$ and $\epsilon_\phi=1.5\%$ (red band);  (ii) $G_\phi=10^{20} G_{\rm F}$ and $\epsilon_\phi=3.0\%$ (light-blue band);  (iii) $G_\phi=10^{21} G_{\rm F}$ and $\epsilon_\phi=1.5\%$ (green band);  (iv) $G_\phi=10^{22} G_{\rm F}$ and $\epsilon_\phi=3.0\%$ (orange band). The data points correspond to the same ones displayed in figure \ref{fig:Peeidata1}. However, the data with the lowest energy corresponds to the expected precision to be attained  by the Darwin experiment in measuring $P_{ee}\pm\Delta P_{ee}$, for which $\Delta P_{ee}$ could be as low as  $\Delta P_{ee}=0.017$ \citep{2020arXiv200603114A}.  The continuous dashed curve corresponds to a 3+1  neutrino model with $G_\phi\approx 0$  and $\epsilon=0\%$.} \label{fig:Peeidata2}
  \end{figure}

\section{Light Dark Matter Impact on Electron Neutrino Spectra}
\label{sec-LDMeneutrinospectrum}

Inside the Sun, the flux variation of neutrinos with different flavors due to matter (including  LDM) is strongly dependent  of the local distributions of electrons and neutrons, but also on the population of dark matter particles in the solar neighbourhood. This new flavor mechanism
(sterile neutrinos and LDM field $\phi$) affects all electron neutrinos produced in the Sun's core. A detailed discussion about the neutrino sources inside the Sun, and their specific solar properties, can be found
in~\citet{2013PhRvD..88d5006L,2017PhRvD..95a5023L}. 
The average survival probability of electron neutrinos for each nuclear reaction in the solar interior, i.e., $P_{e,i} (E,\phi) $ is computed by
\begin{eqnarray} 
P_{ee,i} (E,\phi) = 
C_i \int_0^{R_\odot} P_{e} (E,\phi,r)S_i (r) 4\pi \rho(r) r^2 dr, 
\label{eq:Pnueej}
\end{eqnarray}  
where  $C_i$ $\left(=\left[\int_0^{R_\odot}S_i (r) 4 \pi \rho (r) r^ 2  \;dr\right]^{-1}\right)$ is a normalization constant and 
 $ S_i (r) $  is the electron neutrino emission function  for the $i$ solar nuclear reaction. $i$ corresponds  to the following solar neutrino sources
 (from the PP chain and CNO cycle nuclear reactions): $pp$, $pep$, $^8$B, $^7$Be, $^{13}$N, $^{15}$O and $^{17}$F.

\medskip\noindent
Moreover, since the survival probabilities $P_{ee,i} (E,\phi) $ (Equation \ref{eq:Pnueej}) are time dependent through $\phi$, these quantities also vary with time. Therefore, the oscillation probability $\langle P_{ee}(E) \rangle$ (Equation \ref{eq:Peet}) is generalized for each specific nuclear reaction $i$:
\begin{eqnarray}
\langle P_{ee,i}(E) \rangle=\int_0^{\tau_\phi} P_{e,i} (E,\phi)
\frac{dt}{\tau_\phi}.
\label{eq:Pnueejt}
\end{eqnarray}

\medskip\noindent
The LDM field $\phi$ can lead to different temporal imprints on the neutrino oscillation measurements. The specific impact depends on the mass of the LDM particle.  In the following, we compute the spectra of neutrinos from any specific nuclear reaction that we know to be essentially independent of the properties surrounding  solar plasma.
Since in the 3+1 neutrino flavor model new processes exist to change the survival probability of electron neutrinos, this will modify the solar neutrino spectra measured on Earth. 
These new processes will alter the conversion rates of $\nu_e$ to other flavors 
 ($\nu_\mu$, $\nu_\tau$  and $\nu_s$) and vice versa.  Accordingly, the electron
neutrino spectrum of the nuclear reaction $i$ inside the core is defined
as $\Phi_{i}$, and
$\Phi_{i\odot}$ is the electron neutrino spectrum arriving on Earth \citep{2018EPJC...78..327L} such that: 
\begin{eqnarray}
\Phi_{i,\odot}(E) = P_{ee,i}(E,\phi)  \Phi_{i}(E) 
\label{eq:Phinueejt}
\end{eqnarray}
where $P_{ee,i}(E)$ is the  average survival probability of electron neutrinos 
for reactions in the solar interior as given by equation \ref{eq:Pnueej}. Equally if we take the time average of equation \ref{eq:Phinueejt}, we obtain the following averaged spectrum for each nuclear reaction $i$:
\begin{eqnarray}
\Phi_{i,\odot}(E) =\langle P_{ee,i}(E) \rangle \Phi_{i}(E), 
\label{eq:Phinueemanejt}
\end{eqnarray}
where $\langle P_{ee,i}(E) \rangle $ is the average survival probability of electron neutrinos  as given by equation \ref{eq:Pnueejt}. 

 \begin{figure}[!t]
\centering 
\includegraphics[scale=0.45]{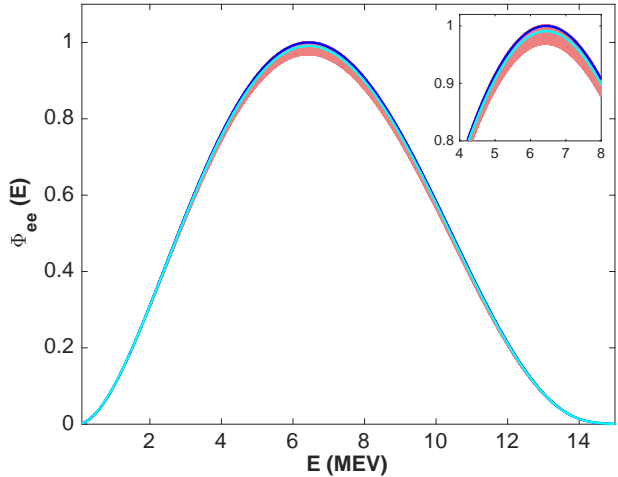}
\caption{The $^8B$ solar electron neutrino spectra measured on Earth detectors. The blue curve corresponds to the standard (undistorted) neutrino spectra, the cyan curve to the averaged (distorted) electron spectra, and the red band corresponds to the ensemble of spectra related with the time-dependent flavor parameters. This  LDM model has a $G_\phi\approx 0$ and an $\epsilon_\phi=3.0\%$. In the calculation of the  $^8$B spectrum we use  a high-Z SSM (see the main text).} \label{fig:PeeMSW1}
\end{figure}

 \begin{figure}[!t]
\centering 
\includegraphics[scale=0.45]{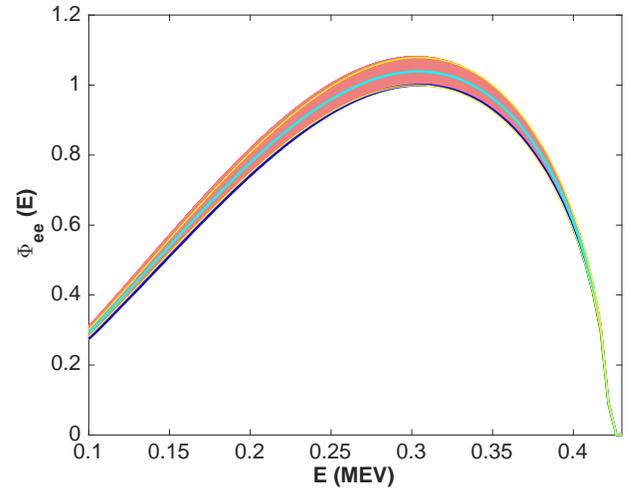}
\caption{The $pp$ solar electron neutrino spectra measured on Earth detectors. The LDM model and  color scheme is the same one as in Figure \ref{fig:PeeMSW1}.  The error bar curves (yellow curves) on the averaged (distorted) electron spectra (cyan curve) corresponds to the expected precision to be attained  by the Darwin experiment	\citep{2020arXiv200603114A}. See Figure \ref{fig:Peeidata2} and main the text.} \label{fig:PeeMSW2}
\end{figure}

\section{The Sun: Light Dark Matter and Sterile Neutrinos} 
\label{sec:sun-neutrinos}

Here,  we will study the impact of the theoretical model presented in the 
previous sections, specifically we compute the survival probability of electron neutrinos
(as given by  equations (\ref{eq:Peesunrad}), (\ref{eq:Peet}), 
(\ref{eq:Pnueej}) and (\ref{eq:Pnueejt})) in the case of a standard 
solar model with low-Z \citep[e.g.,][]{2013MNRAS.435.2109L,2020MNRAS.498.1992C}. 

\medskip\noindent
In the parameterization for the 3+1 neutrino flavor oscillation model,
we opt to adopt the recent values obtained in the data analysis 
of the standard three-neutrino flavor oscillation model obtained by \citet{2020arXiv200611237D}, and for the sterile neutrino additional fiducial parameters we used the values obtained by  \citet{2015JPhG...43c3001G}.  Accordingly, for a parameterisation with  a  normal ordering of neutrino masses, the mass-square difference and the mixing angles have the following values:	
$\Delta m^2_{21}= 7.50^{+0.22}_{-0.20}\times 10^{-5}{\rm eV^2}$, 
$\sin^2{\theta_{12}}=0.318\pm 0.016 $,
and  $\sin^2{\theta_{13}}=0.02250^{+0.00055}_{-0.00078}$.
Although, $\Delta m^2_{31}= 2.56^{+0.0003}_{-0.0004}\times 10^{-3}{\rm eV^2}$ and $\sin^2{\theta_{23}}=0.56^{+0.016}_{-0.022}$, we mention them here for reference
\citep{2020arXiv200611237D}. These new parameters are consistent with  previous estimations \citep{2019JHEP...01..106E, 2016NuPhB.908..199G}.  
For the sterile neutrino, we  choose the following fiducial values for the mass-square difference and mixing angles 
~\citep{2015JPhG...43c3001G,2016JPhG...43c3001G,2017JCAP...07..021C}:
$\Delta m^2_{41}=1.6 \ {\rm eV^2}$,  $\sin^2{\theta_{14}}=0.027$, $\sin^2{\theta_{42}}=0.014$ and the other mixing angle for the sterile
neutrinos are fixed to zero. Moreover,   we assume that all phases ($\delta_{13,14,34}$) 
and other angles related to the sterile neutrino are equal to zero.

\medskip\noindent
The present-day internal structure of the Sun corresponds to 
an up-to-date standard solar model (SSM) that has a better 
agreement with neutrino fluxes and  helioseismic data sets.  
This solar model was obtained from a one-dimensional stellar evolution
code allowed to evolve in time until the present-day solar age, $4.57$ Gyr, having been calibrated to the values of luminosity and effective temperature of the present Sun, of $ 3.8418 \times 10^{33}$ erg s$^{-1}$ and $ 5777$ K, respectively, as well as the observed abundance ratio at the Sun's surface: ($Z_\text{s}/X_\text{s}$)$_{\odot}=0.0181$, where $Z_s$ and $X_s$ are the metal and hydrogen abundances at the surface of the star \citep{1993ApJ...408..347T,1995RvMP...67..781B,2006ApJS..165..400B}. 
This stellar model was computed with  the release version 12115 of the stellar evolution code MESA \citep{2011ApJS..192....3P,2019ApJS..243...10P}. 
The details about the physics of this standard solar model in which we use the AGSS09 (low-Z) solar abundances \citep{2009ARA&A..47..481A}   are  described in \citet{2013MNRAS.435.2109L} and \citet{2020MNRAS.498.1992C}.  

\medskip\noindent
Figures \ref{fig:Petime1}  and \ref{fig:Petime2} show the impact of the time-dependent  mass-square difference (Equation \ref{eq:deltamphi}) and  mixing angles  (Equation \ref{eq:thetaphi}) on the averaged electron survival probability 
$\langle P_{ee}(E) \rangle$ (Equation \ref{eq:Peet}) for which the LDM field $\phi$ has a fixed amplitude (Equation \ref{eq:etaphi}): $\epsilon_\phi=0$ or $\epsilon_\phi=1.5\%$. 
We also show LDM models for which the sterile neutrino couples to $\phi$ with strength  $G_\phi$.

\medskip\noindent
The overall shape of the curve $\langle P_{ee}(E) \rangle$  depends on $G_\phi$ times $ n_\phi$ in the potential $V_{\nu_s\phi}$ or the ratio $G_\phi/m_\phi$ as previously mentioned. 
For instance, in a LDM model in which we fix  $m_\phi =10^{-9}\; {\rm eV }$ (or $n_\phi$), an increase of $G_\phi$ from $10^{20}$ to $10^{22}\;G_{\rm F}$ leads $\langle P_{ee}(E) \rangle$ to vary significantly, as shown in Figure \ref{fig:Petime1}. As expected this change in $\langle P_{ee}(E) \rangle$  is more pronounced for high energy neutrinos where the MSW effect is more significant. If we choose higher values of $m_\phi $ the  results will somehow be similar (see Figure \ref{fig:Petime1}). 
 
 \medskip\noindent
Evidently, for an LDM model in which $m_\phi$ decreases by a certain amount ($n_\phi=\rho_{\phi}/m_\phi$), the constancy of $G_\phi/ m_\phi$ in $V_{\nu_s\phi}$ implies that $G_\phi$ can increase by the same order of magnitude to obtain the same MSW effect on the $\langle P_{ee}(E) \rangle$ curve (see Figure \ref{fig:Petime1}). For instance, an LDM model with  $m_\phi=10^{-23}\;{\rm eV}$ and $G_\phi = 10^{7} G_{\rm F}$ will have  $\langle P_{ee}(E) \rangle$  identical to an LDM model with  $m_\phi=10^{-9}\;{\rm eV}$ and $G_\phi = 10^{21} G_{\rm F}$ , since in both  LDM models we have the same ratio: $G_\phi/m_\phi\approx 10^{30}$. The same argument explains the reason why the coupling constant between sterile neutrinos and more massive dark matter particles is much smaller in those models than in the present study. For instance, this is the case for particles captured from the dark matter halo by the Sun.  Since over time, the star accreted a significant amount of dark matter \citep[e.g.,][]{2018ApJ...869..112L}, for these models 
$G_\phi$  is significantly smaller than the value found for the present study.  

\medskip\noindent
The most important feature of such a class of LDM models is the time dependence of the dark matter field $\phi(t)$ and its imprint in the flavor oscillation parameters' mass-square differences (Equation \ref{eq:deltamphi})  and mixing angles (Equation \ref{eq:thetaphi}).  
As predicted by equation \ref{eq:Peesunrad}, there are many $P_{ee}(E,\phi)$  with near similar behavior. Figure  \ref{fig:Petime1} shows  an ensemble of time-dependent $ P_{ee}(E,\phi) $  as a pink band.  The difference between curves  relates to the dependence of the oscillation parameters on time. In  this LDM model it is assumed there is a very negligible interaction between sterile neutrinos and $\phi$ (for which $G_\phi \approx 0$).   The figure also  shows $\langle P_{ee}(E) \rangle$  (red curve)  the time-averaged of the ensemble of $P_{ee}(E,\phi)$ curves that we compute using equation  \ref{eq:Peet}. 

\medskip\noindent
Although there are several parameters that contribute for the time variability of  $P_{ee}(E,\phi)$  (Equation \ref{eq:Peesunrad}), the main contributions come from $\theta_{21}$ and $\theta_{13}$. The variability  related   $\theta_{24}$ is relevant for the high energies. We notice that the contributions coming from $\Delta m^2_{21}$ and $\theta_{32}$ are much smaller than all the parameters mentioned above.
The amplitude of the $P_{ee}(E,\phi)$  pink band  is defined by the value of $\epsilon_\phi $ for which we adopt the fiducial value of $\epsilon_\phi =1.5\%$.
It is worth pointing out that the $P_{ee}(E,\phi)$ band is much larger for low-energy than for higher-energy values. Moreover, the averaged value of this ensemble given by $\langle P_{ee}(E) \rangle$ is identical to $\langle P_{ee}(E) \rangle$ with $\epsilon_\phi\approx 0$.  Figure \ref{fig:Petime2} shows the variability of $P_{ee}(E,\phi)$  for  a LDM with different $G_\phi$ values.  These results are identical to the model in figure  \ref{fig:Petime1}. Nevertheless, the LDM model with the largest $G_\phi$ has (an orange) band with a smaller amplitude around $\langle P_{ee}(E) \rangle$. Once again, the band thickness decreases for neutrinos with higher energy for all these models. 

\medskip\noindent
Figures \ref{fig:Peeidata1} and \ref{fig:Peeidata2} compare our predictions with current solar neutrino data \citep[e.g.,][]{2020arXiv200603114A}. These figures show that LDM models with relatively low values of $\epsilon_\phi$ and   $G_\phi$ are compatible  overall  with current solar neutrino data coming from Borexino, Super-Kamiokande, and SNO. 
Clearly, this analysis has also shown that the precision of our current solar neutrino experiments is not able to distinguish between some of these LDM models. Nevertheless, it is already possible to put some constraints on these LDM models. For instance, we found that LDM models with $G_\phi\approx 0$ must have a $\epsilon_\phi$ smaller than $3\%$ to be consistent with all data, including $pp$ measurements of the Borexino detector \citep{2020PhRvD.101f2001A} (see Figure \ref{fig:Peeidata1}); and  any LDM models must have a ratio $G_\phi/m_\phi$ smaller than $10^{30}$, otherwise they become inconsistent with $pp$ and $^7$Be measurements for several solar detectors \citep{2020PhRvD.101f2001A,2019PhRvD.100h2004A,2018Natur.562..505B,2010PhRvD..82c3006B} (see figure \ref{fig:Peeidata2}). Figure \ref{fig:Peeidata2} shows a LDM model with a $m_\phi= 10^{-9}\; {\rm eV}$ and  $G_\phi=10^{22}\; G_{\rm F}$ with a ratio $G_\phi/m_\phi$ of the order of $10^{31}\;G_{\rm F}$ (see Figure \ref{fig:Peeidata2});  This ratio is one order of magnitude larger than the critical $G_\phi/m_\phi$ value of $10^{30}$ discussed in the previous section. Figure \ref{fig:Peeidata2} also shows that the variability related with time dependence on $P_{ee}(E,\phi)$ decreases for large values of $G_\phi$.

\medskip\noindent
There is another important effect that also contributes to the time variability of $P_{ee}(E,\phi)$. 
The PP chain and CNO cycle, nuclear reactions occur at different distances from the center of the Sun and each nuclear reaction emits neutrinos in a well-defined energy range. As a consequence,  the electron neutrinos produced in each specific nuclear reaction will be affected differently by the MSW effect. As such, this effect will also contribute to the overall variability of electron neutrinos  $P_{ee,i}(E,\phi)$ (see Equation \ref{eq:Pnueej}) and their time-averaged  
$\langle P_{ee,i}(E,\phi)\rangle $ (see Equation \ref{eq:Pnueejt}). 

\medskip\noindent
The time-dependent electron neutrino survival probability will have a significant impact on the neutrino spectra of the different nuclear reactions.  Accordingly,  figures \ref{fig:PeeMSW1} and \ref{fig:PeeMSW2}  show the spectra correspond to two neutrino types: $pp$ and $^8B$ neutrinos.   An essential difference between these two spectra relates to the thickness of the $\epsilon_\phi$ band for a fixed value since thickness decreases with neutrino energy. Therefore the $\epsilon_\phi$ band is more significant for a $pp$ spectrum than for a  $^8B$ neutrino spectrum.  This is an effect identical to the one discussed previously for the $P_{ee}(E,\phi) $ functions. Therefore, the measurement of solar neutrino fluxes and solar neutrino spectrum in the energy range  below $0.2\; MeV$ will provide the strongest constraint for such a class of dark matter models.  Figure \ref{fig:PeeMSW1} and \ref{fig:PeeMSW2}  show the spectra of $^8$B and $pp$, if we assume the  precision expected  to be attained  by the Darwin experiment \citep{2020arXiv200603114A}.  
Figure \ref{fig:PeeMSW2} also shows the precision expected for the  Darwin experiment.

\section{Conclusion}
\label{sec-DC}

This article focuses on the impact of LDM on solar neutrino fluxes, spectra, and survival probabilities of electron neutrinos, specifically a dark matter model made of two particles: a sterile neutrino and  an LDM particle.  In particular, we describe how the 3+1 neutrino flavor model is affected by this type of LDM particles, with an emphasis on how the LDM affects the Wolfenstein potentials.   We also study how the dark matter models affect the survival probability functions of electron neutrinos  related to the different nuclear reactions occurring in the solar interior, and we compute the spectra of two  relevant solar neutrino sources: $pp$ and $^8B$ neutrino nuclear reactions.   

\medskip\noindent
By studying a large range of dark matter particle masses (from  $10^{-9}$ to $10^{-23}\;{\rm eV}$) we found that depending on the mass of these LDM particles and the value of the generalized Fermi constant, the shape of electron neutrino survival probability and their spectra can vary with time. 
We establish that for LDM particles with low masses (low-frequency regime), the solar neutrino detectors can observe the electron neutrino survival probability changing with time. Conversely,  for dark matter particles with higher masses (high-frequency regime), this impact can be determined by measuring the time-averaged electron neutrino survival probability. 

\medskip\noindent
It was possible to establish, using data of current solar neutrino measurements, that those models with a $G_\phi/m_\phi $ ratio smaller than  $10^{30}\; G_{\rm F} eV^{-1}$ agree with current solar neutrino data from the Borexino, SNO and Super-Kamiokande detectors.  We also found that for models with a near-zero constant, the time-variability amplitude must be smaller than 3\%. Such a constraint is equivalent to the condition  ${g_\phi\sqrt{2\rho^{\odot}_{\rm DM}}}/
{(m_\phi m_{\nu})}\; \left({\Omega_{\phi}}/{\Omega_{\rm DM}}\right)^{1/2}\le 0.03$.

\medskip\noindent
Finally, we also found that the precision expected in the measurements to be made by the Darwin detector will allow us to put powerful constraints to this class of models.

\begin{acknowledgments}
The author is grateful to the  MESA  team for having made their code publicly available. The author also thanks the anonymous referee for the revision of the manuscript.
I.L. thanks the Funda\c c\~ao para a Ci\^encia e Tecnologia (FCT), Portugal, for the financial support to the Center for Astrophysics and Gravitation (CENTRA/IST/ULisboa) 
through the Grant Project~No.~UIDB/00099/2020  and grant No. PTDC/FIS-AST/28920/2017.
\end{acknowledgments}

\bibliographystyle{yahapj}
\begin{flushright}
	
\end{flushright}


\end{document}